\documentclass[conference]{IEEEtran}
\usepackage[latin9]{inputenc}
\usepackage{color}
\usepackage{amsmath}
\usepackage{amssymb}
\usepackage{graphicx}
\usepackage{fancyhdr}
 
\makeatletter
\IEEEoverridecommandlockouts
\usepackage{cite}
\usepackage{amsfonts}\usepackage{algorithmic}
\usepackage{subfigure}
\usepackage{textcomp}
\usepackage{xcolor}
\def\BibTeX{{\rm B\kern-.05em{\sc i\kern-.025em b}\kern-.08em
    T\kern-.1667em\lower.7ex\hbox{E}\kern-.125emX}}

\makeatother

\begin{document}

\title{Degenerate Distributed Feedback Photonic Structure with Double Grating
Exhibiting Degenerate Band Edge 
}

\author{\IEEEauthorblockN{Tarek Mealy, and Filippo Capolino} \IEEEauthorblockA{\textit{Department of Electrical Engineering and Computer Science,
University of California, Irvine, CA 92697 USA} \\
tmealy@uci.edu and f.capolino@uci.edu}}
\maketitle
\thispagestyle{fancy}
\begin{abstract}
We propose a degenerate version of the Bragg condition, associated
to a degenerate band edge (DBE). A standard Bragg condition can be
implemented using a periodic grating that operates at the regular
band edge. The structure we propose to realize a DBE is made of two
stacked identical gratings that form two coupled periodic waveguides
with broken mirror symmetry. The occurrence of the DBE is verified
by using both an eigenmode solver and the calculations of scattering
parameters, using full-wave simulations. The proposed structure is
a good candidate to conceive a degenerate distributed feedback lasers
operating at the DBE.
\end{abstract}

\begin{IEEEkeywords}
Exceptional point of degeneracy; Double grating waveguide; Degenerate
band edge laser; Distributed feedback laser. 
\end{IEEEkeywords}

\IEEEpeerreviewmaketitle{}

\begin{figure}
\begin{centering}
\centering \subfigure[]{\includegraphics[width=0.7\columnwidth]{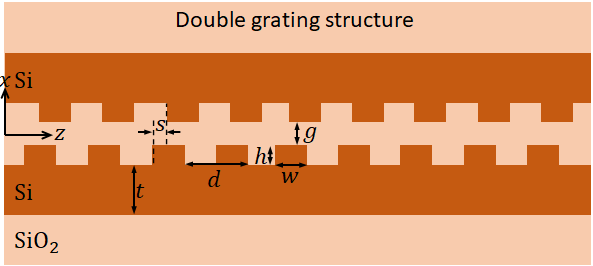}\label{Fig:Double_Grating}} 
\par\end{centering}
\begin{centering}
\centering\subfigure[]{\includegraphics[width=0.8\columnwidth]{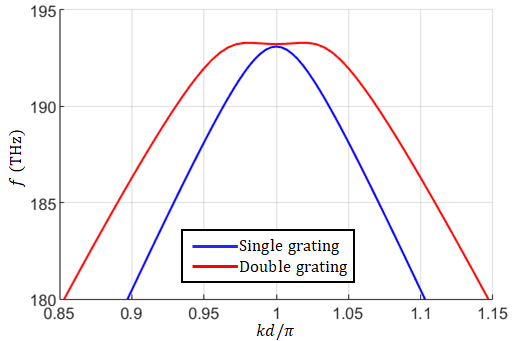}\label{Fig:Double_Grating_Disp}} 
\par\end{centering}
\begin{centering}
\centering\subfigure[]{\includegraphics[width=0.7\columnwidth]{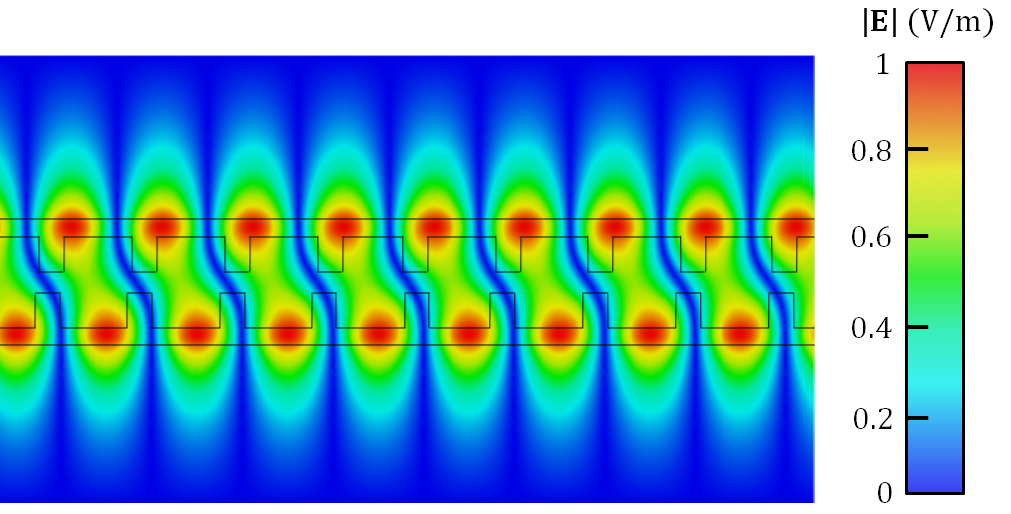}\label{FigField}} 
\par\end{centering}
\centering{}\caption{(a) Degenerate distributed feedback photonic cavity with double grating,
based on a DBE, obtained by braking mirror symmetry. (b) Dispersion
diagram of the eigenmodes in the double grating (a) when the geometry
is optimized to exhibit a DBE (red curve). We also show the dispersion
of modes in a single grating (blue curve). Note the much flatter dispersion
of the mode in the double graing (red curve). (c) Electric field distribution
of the DBE mode calculated at $kd=\pi$. }
\end{figure}

We present a degenerate version of the Bragg condition in a double
grating, leading to a 4th order degeneracy, namely a degenerate band
edge (DBE), where four eigenmodes coalesce forming a single degenerate
eigenmode. Compared to previous studies \cite{figotin2005gigantic,figotin2011slow,Sukhorukov,Gutman,burr2013degenerate,wood2015degenerate,nada2018giant}
devoted to the fundamental aspects of the DBE, here we build on those
concepts and specifically show a realistic structure that supports
a DBE, based on two coupled dielectric layers with periodic gratings,
shown in Fig. \ref{Fig:Double_Grating}. The standard Bragg condition
$kd=m\pi$ for a periodic waveguide, where $k$ is the Bloch wavenumber,
$d$ is the waveguide period, and $m$ is an integer, is associated
with the regular band edge (RBE) \cite{kogelnik1972coupled,ghafouri2003distributed}.
A Bragg condition implies that two modes with power flux in opposite
directions merge at a band edge to form a degenerate mode with vanishing
group velocity. The Bragg condition is used in distributed feedback
(DFB) lasers so that light is reflected in a distributed fashion within
a cavity rather than only at the end mirrors \cite{kogelnik1972coupled,wang1974principles,degnan1976waveguide,ghafouri2003distributed}.
The Bragg condition is implemented through a periodic grating that
acts as a dielectric waveguide. DFB lasers require the presence of
an active medium in addition to satisfying the Bragg condition $kd=m\pi$
to form a self sustained resonant mode that keeps oscillating \cite{ghafouri2003distributed}.
The ``degenerate DFB'' concept presented here could improve laser
properties.

The DBE is a particular kind of exceptional point of degeneracy (EPD)
of order $4$ in a waveguide without loss and gain, where $4$ eigenmodes
coalesce in their wavenumbers and polarization states. The concept
of degeneracy of the eigenvalues and eigenvectors and their perturbation
was discussed in \cite{vishik1960solution,lancaster1964eigenvalues,kato2013perturbation,seyranian1993sensitivity},
and more recently also in \cite{figotin2005gigantic,figotin2011slow,gutman2012slow,nada2018giant,abdelshafy2018exceptional,Mealy2020general}.
Here, we use those concepts to show that the double grating in Fig.
\ref{Fig:Double_Grating} exhibits a DBE (i.e., an EPD of order four)
by engineering the structure's geometrical dimensions. At a DBE the
dispersion diagram is locally represented as $\omega_{d}-\omega=h\left(k-k_{d}\right)^{4}$,
where $\omega_{d}$ and $k_{d}$ are the DBE angular frequency and
wavenumber, respectively, and the parameter $h$ describes the flatness
of the dispersion curve. Therefore, because of the very flat dispersion,
not only the degenerate mode has a vanishing group velocity $v_{g}=0$,
but also its second and third derivatives vanish. Photonic devices
operating at the DBE have the potential to exhibit better performance
than those operating at the RBE \cite{Sukhorukov,Gutman,burr2013degenerate,wood2015degenerate,veysi2018degenerate,nada2018giant,abdelshafy2020distributed,oshmarin2021experimental}.

The waveguide with double grating in Fig. \ref{Fig:Double_Grating}
consists of a standard grating over a dielectric layer, coupled to
another grating obtained by a mirror operation in the $x$ direction,
followed by a translation $s$ along $z$. The shift $s$ breaks mirror
symmetry, because mirror symmetry prohibits the existence of a DBE
in two coupled waveguides due to the existence of even and odd modes
that are always decoupled from each other. The double grating in Fig.
\ref{Fig:Double_Grating} is made of two silicon on insulator (SOI)
waveguides with silicon refractive index $n_{\mathrm{Si}}=3.45$,
and with a cladding material of silicon dioxide with refractive index
$n_{\mathrm{SiO2}}=1.97$.

\begin{figure}
\begin{centering}
\centering \includegraphics[width=0.9\columnwidth]{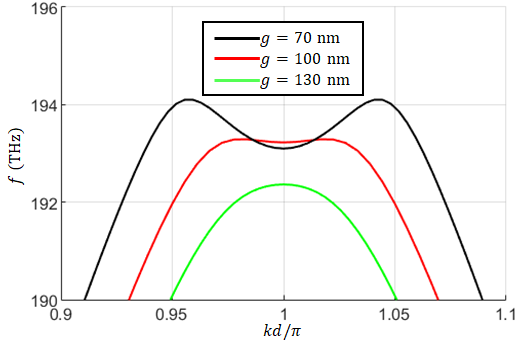}\label{Fig:Sweep}
\par\end{centering}
\centering{}\caption{Dispersion diagram of modes in the double grating in Fig. \ref{Fig:Double_Grating}
for three distinct gap values. The change of the dispersion curve
from a split band edge ($g=70$ nm) to regular band edge ($g=130$
nm) implies the existence of DBE using a value of $g$ between 70
nm and 130 nm. Indeed, the figure shows that the DBE occurs when $g=100$
nm (red curve).}
\end{figure}

The modal dispersion relation in Fig. \ref{Fig:Sweep} is obtained
using the full-wave eigenmode solver implemented in CST Studio Suite.
We used periodic boundaries at two virtual cross sections at a distance
of a period $d.$ We look for modes polarized along $y$, hence we
use perfect electric conductor (PEC) walls at planes with constant
\textit{y} to account for field invariance in the $y$ direction.
The computational domain is truncated by PEC also at $x=\pm12t$,
where $t$ is the dielectric layer thickness as illustrated in Fig.
\ref{Fig:Double_Grating}, and it is sufficiently large not to affect
the field in the grating area, to ensure that the studied modes decay
away from the waveguide in the positive and negative $x$ directions.

The dielectric waveguide has thickness $t=67$ nm, the gratings has
period $d=$335 nm, width $w=87$ nm, and height $h=121$ nm. The
rest of parameters, the coupling gap $g$ and the shift between the
waveguides $s$, are left to be optimized to achieve a DBE. An optimization
procedure based on having a DBE in the dispersion relation leads to
$g=100$ nm and $s=20$ nm. The dispersion relation obtained from
the eignmode solver for the optimized unit cell in the Fig. \ref{Fig:Double_Grating_Disp}
shows a DBE at around 193 THz (red curve). The flatness of the DBE
is compared to that of the the dispersion of modes in the single grating
structure (blue curve) with the same waveguide and grating parameters
except for the grating height that is now 42 nm in order to have a
regular band edge more or less at the DBE frequency of 193 THz. It
is clear from the figure that the dispersion relation for the double
grating structure (red curve) is flatter than that of single grating
structure (blue curve). The field distribution of the DBE mode at
$k=\pi/d$ is shown in Fig. \ref{FigField}.

Figure \ref{Fig:Sweep} shows the dispersion relation of three distinct
designs of the double grating structure in Fig. \ref{Fig:Double_Grating}
based on three distinct values of the coupling gap $g$, the rest
of the dimensions are the same as the ones mentioned for the case
above. Note that for $g=70$ nm the dispersion shows a split band
edge \cite{noh2010giant}, with four modes around $f=193$ THz (black
curve), which coalesce when using $g=100$ nm. For larger values of
$g$, the green curve shows a RBE. Indeed, when a parameter is swept
(e.g., $g$) and one finds both a RBE and a split band edge, there
is a proper intermediate value (here, $g=100$ nm) that leads to a
DBE \cite{noh2010giant}.

\begin{figure}
\begin{centering}
\centering \subfigure[]{\includegraphics[width=1\columnwidth]{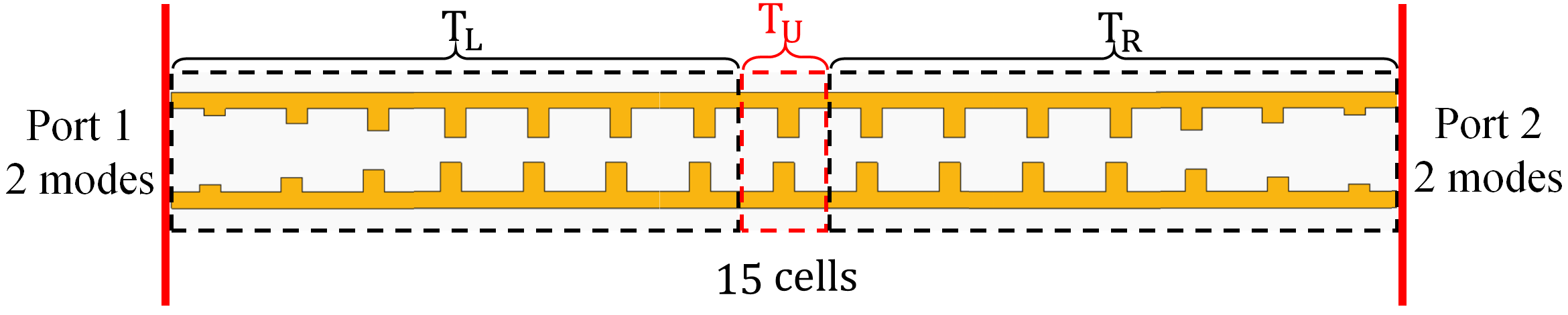}\label{Fig:Str1}}
\par\end{centering}
\begin{centering}
\centering\subfigure[]{\includegraphics[width=1\columnwidth]{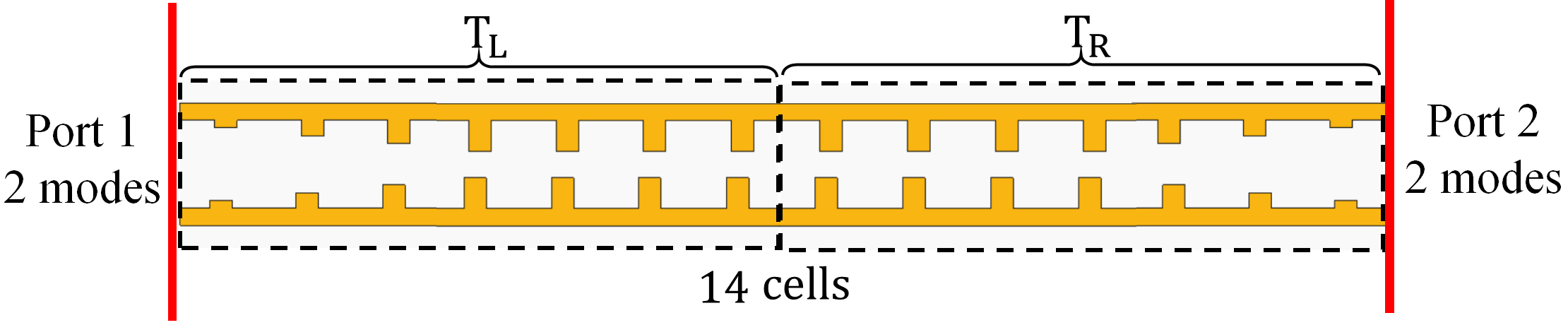}\label{Fig:Str2}}
\par\end{centering}
\begin{centering}
\centering\subfigure[]{\includegraphics[width=0.99\columnwidth]{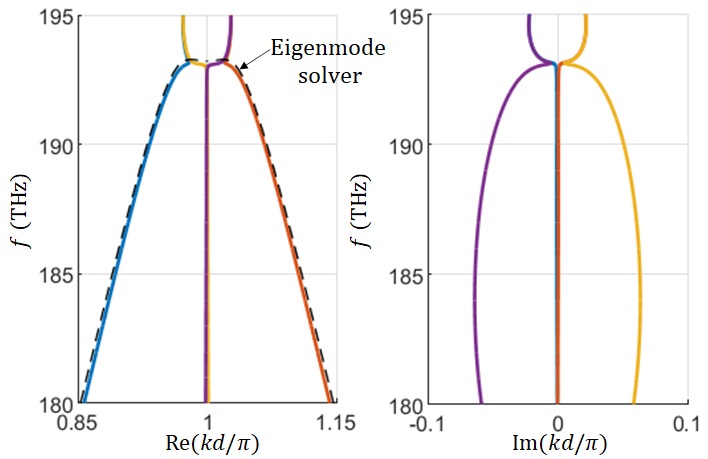}\label{Fig:Disp_Complex}}
\par\end{centering}
\centering{}\caption{(a) Finite-length structure used to estimate the transfer matrix \textcolor{black}{$\underline{\mathbf{T}}_{\mathrm{U}}$}
of one unit cell in the middle. (b) Structure used to de-embed the
effect of the two transition regions connected to the left and right
sides of the unit cell in (a). (c) Dispersion diagram showing four
complex-valued wavenumber versus frequency illustrating the coalescence
of four wavenumbers at the DBE point $k_{d}d=\pi$ , which is the
degenerate Bragg condition. The dispersion is obtained based on the
transfer matrix \textcolor{black}{$\underline{\mathbf{T}}_{\mathrm{U}}$}
found by dembedding the structure in (b) from (a). The black-dashed
curve shows the dispersion of the purely real branch obtained from
the eigenmode solver, already plotted in Fig. Fig. \ref{Fig:Double_Grating_Disp},
for validation.}
\end{figure}

\textcolor{black}{We also verify the existence of the DBE in the proposed
double grating structure by finding the scattering matrix using frequency-domain
full-wave simulations. Open (radiation) boundaries are used in this
simulation and we use waveports as shown in Fig. \ref{Fig:Str1} where
two modes (even and odd) are excitable on each port. Note that simulating
one unit cell would not lead to an accurate transfer matrix because
it would not fully account for the structure periodicity since a waveport
supports a waveguide mode and not a Bloch mode. Therefore, in the
simulation we include two segments, on the left and right sides of
the unit cell, that act as adiabatic transitions between the dielectric
waveguide mode and the grating, as shown in Fig. \ref{Fig:Str1}.
The transition is a chirped grating with height steps of 30 nm, until
we reach the whole grating height $h=120$ nm. However our goal is
to find the transfer matrix $\underline{\mathbf{T}}_{\mathrm{U}}$
only. Therefore, we also simulated another structure with two chirped
gratings as shown in Fig. \ref{Fig:Str2}, which is the same structure
as in Fig. \ref{Fig:Str1} but without the unit cell at the center.
The structure in Fig. \ref{Fig:Str2} is used to }deembed\textcolor{black}{{}
the effect of the two transition segments from the results relative
to Fig. \ref{Fig:Str1}, similarly to what was done in \cite{apaydin2012experimental,nada2020frozen}.}

Using even and odd modes in the ports shown in Fig.\textcolor{black}{{}
\ref{Fig:Str1}} and \textcolor{black}{\ref{Fig:Str2}}, and defining
their associated incident and reflected waves

\begin{equation}
\begin{array}{c}
\mathbf{b}=\left[\begin{array}{cccc}
b_{\mathrm{e}1}, & b_{\mathrm{o}1}, & b_{\mathrm{e}2}, & b_{\mathrm{o}2}\end{array}\right]^{T},\\
\mathbf{a}=\left[\begin{array}{cccc}
a_{\mathrm{e}1}, & a_{\mathrm{o}1}, & a_{\mathrm{e}2}, & a_{\mathrm{o}2}\end{array}\right]^{T}.
\end{array}
\end{equation}
such that $\mathbf{b=\underline{\mathbf{S}}a}$. Using full-wave simulations,
we find the $4\times4$ scattering matrix expressed as

\begin{equation}
\underline{\mathbf{S}}=\left[\begin{array}{cc}
\mathbf{S}_{11} & \mathbf{S}_{12}\\
\mathbf{S}_{21} & \mathbf{S}_{22}
\end{array}\right].\label{eq:S_matr_DBE_Douvble}
\end{equation}
 It is convenient to define the state vectors $\mathbf{\mathbf{\Psi}}_{1}=\left[\begin{array}{cccc}
a_{\mathrm{e}1}, & a_{\mathrm{o}1}, & b_{\mathrm{e}1}, & b_{\mathrm{o}1}\end{array}\right]^{T}$ and $\mathbf{\mathbf{\Psi}}_{2}=\left[\begin{array}{cccc}
b_{\mathrm{e}2}, & b_{\mathrm{o}2}, & a_{\mathrm{e}1}, & a_{\mathrm{o}1}\end{array}\right]^{T}$, relative to port 1 and port 2, respectively. They are defined in
way that the first two elements represent the amplitudes of waves
propagating in the positive $z$ direction, whereas the second two
elements represent the amplitudes of waves propagating in the negative
$z$ direction. As a consequence, the transfer matrix $\underline{\mathbf{T}}$
relating them, $\mathbf{\mathbf{\Psi}}_{2}=\underline{\mathbf{T}}\mathbf{\mathbf{\Psi}}_{1}$,
is found as \cite{nada2017theory}

\begin{equation}
\underline{\mathbf{T}}=\left[\begin{array}{cc}
\mathbf{S}_{21}-\mathbf{S}_{22}\mathbf{S}_{12}^{-1}\mathbf{S}_{11} & \mathbf{S}_{22}\mathbf{S}_{12}^{-1}\\
-\mathbf{S}_{12}^{-1}\mathbf{S}_{11} & \mathbf{S}_{12}^{-1}
\end{array}\right].\label{eq:DBE_Double_Trans}
\end{equation}

\textcolor{black}{The T-matrices of the two structures shown in Figs.
\ref{Fig:Str1} and \ref{Fig:Str2} are $\underline{\mathbf{T}}_{\mathrm{15}}=\underline{\mathbf{T}}_{\mathrm{R}}\underline{\mathbf{T}}_{\mathrm{U}}\underline{\mathbf{T}}_{\mathrm{L}}$
and $\underline{\mathbf{T}}_{\mathrm{14}}=\underline{\mathbf{T}}_{\mathrm{R}}\underline{\mathbf{T}}_{\mathrm{L}}$,
respectively, where $\underline{\mathbf{T}}_{\mathrm{R}}$ and $\underline{\mathbf{T}}_{\mathrm{L}}$
are the T-matrices of the two chirped grating structures. The chirped
structure represented by $\underline{\mathbf{T}}_{\mathrm{R}}$ is
a flipped version, in the $z$ direction, of the other one represented
by $\underline{\mathbf{T}}_{\mathrm{L}}$, and therefore they satisfy
$\underline{\mathbf{T}}_{\mathrm{R}}=\mathbf{F}\underline{\mathbf{T}}_{\mathrm{L}}^{-1}\mathbf{F}^{-1}$,
if $\underline{\mathbf{T}}_{L}$ is not singular, where $\mathbf{F}$
is a transformation matrix that flips the direction of wave propagation,
made by two $\mathbf{0}$ in the diagonal blocks and two identity
matrices in the codiagonal blocks.}

\textcolor{black}{Hence, we calculate the new T-matrix $\underline{\mathbf{T}}_{\mathrm{n}}=\underline{\mathbf{T}}_{\mathrm{15}}\underline{\mathbf{T}}_{\mathrm{14}}^{-1}=\underline{\mathbf{T}}_{\mathrm{R}}\underline{\mathbf{T}}_{\mathrm{U}}\underline{\mathbf{T}}_{\mathrm{R}}^{-1}$
whose eigenvalues are the same of those of the unit-cell T-matrix
$\underline{\mathbf{T}}_{\mathrm{U}}$, similarly to what is done
in \cite{apaydin2012experimental,nada2020frozen}. The T-matrices
$\underline{\mathbf{T}}_{\mathrm{15}}$ and $\underline{\mathbf{T}}_{\mathrm{14}}$
are obtained by transformation of the scattering matrices associated
to the 4-port structures in Figs. \ref{Fig:Str1} and Fig. \ref{Fig:Str2}
based on\ref{eq:DBE_Double_Trans}.}

According to Floquet-Bloch theory, we look for periodic solutions
of the state vector as $e^{-jkd}$ where $k$ is the Floquet-Bloch
complex wavenumber, that satisfy $\mathbf{\boldsymbol{\Psi}}'=\lambda\mathbf{\boldsymbol{\Psi}}$,
with $\lambda\equiv e^{-jkd}$, where $\mathbf{\boldsymbol{\Psi}}$
and $\mathbf{\boldsymbol{\Psi}}'=\mathbf{\underline{T}}_{\mathrm{U}}\mathbf{\boldsymbol{\Psi}}$
are the state vectors at the input and output of a unit cell. The
eigenvalue problem is then formulated as

\begin{equation}
\mathbf{\underline{T}}_{\mathrm{U}}\mathbf{\boldsymbol{\Psi}}=\lambda\mathbf{\boldsymbol{\Psi}},\label{eq:Eig_prob}
\end{equation}
where the eigenvalues $\lambda_{n}\equiv e^{-jk_{n}d}$, with $n=1,2,3,4$,
are obtained by solving the dispersion characteristic equation $D(k,\omega)\equiv\mathrm{det}(\mathbf{\underline{T}}_{\mathrm{U}}-\lambda\mathbf{\underline{1}})$,
with $\mathbf{\underline{1}}$ being the $4\times4$ identity matrix. 

We show in Fig. \ref{Fig:Disp_Complex} the Floquet-Bloch complex-valued
wavenumber dispersion diagram of the four eigenmodes of the structure
in Fig. \ref{Fig:Double_Grating}. The dispersion diagram confirms
the existence of the DBE at a frequency $f\approx193$ THz, where
four branches coalesce. The black-dashed curve is the dispersion diagram
with a purely real wavenumber obtained via the eigenmode solver using
phase-shift boundary conditions, shown here for comparison purposes.
Indeed, a good matching is found for the curve with purely real wavenumber,
which validates the method described above based in the dembedding
procedure, that confirms the existence of the DBE.

In conclusion, we have proposed a photonic structure with double grating
that exhibits a DBE in its dispersion diagram where a degenerate Bragg
condition occurs at $kd=\pi$. We have demonstrated the occurrence
of the DBE using two methods: (i) by finding the dispersion relation
of modes with purely real wavenumbers using an eigenmode solver and
observing the very flat dispersion, and (ii) by finding the four complex-valued
wavenumbers from the scattering parameters by using full-wave simulations.
The degenerate Bragg condition we have found in the double grating
structure can be used to conceive a ``degenerate distributed feedback
laser'', that may lead to interesting properties, even augmented
when compared to a DFB laser. The first DBE laser analysis was carried
out in \cite{veysi2018degenerate} using a transmission line method
for an ideal structure. Interesting properties were found in that
ideal DBE structure like stability of the lasing oscillation frequency
when varying the load, and lower threshold than the one in the RBE
laser, even in the case when the RBE cavity has the same quality factor
of the DBE cavity. Future studies shall be devoted to study the cavity
effect and the lasing properties of the double grating with degenerate
Bragg condition presented in this letter.

\section*{Acknowledgment}

This material is based on work supported by the Air Force Office of
Scientific Research award number FA9550-18-1-0355, and by the National
Science Foundation under award NSF ECCS-1711975. The authors are thankful
to DS SIMULIA for providing CST Studio Suite that was instrumental
in this study.

\bibliographystyle{ieeetr}
\bibliography{My_Ref}

\end{document}